\pdfoutput=1
\documentclass[11pt, journal, onecolumn, oneside]{IEEEtran}

\usepackage[utf8]{inputenc}
\usepackage[T1]{fontenc}
\usepackage{cite}
\usepackage{amsmath,amssymb,amsfonts}
\usepackage{algorithmic}
\usepackage{algorithm}
\usepackage{graphicx}
\usepackage{textcomp}
\usepackage{xcolor}
\usepackage{booktabs}
\usepackage{hyperref}

\begin{document}

\title{Overcoming the BCI Calibration Bottleneck: A Clinically-Grounded Architecture using Riemannian Alignment and Stochastic Weight Averaging}

\author{Immanuvel Prathap Sagayaraju \\
\textit{MSc Artificial Intelligence in Medicine}\\
\textit{University of Bern}\\
\textit{Bern, Switzerland}\\
\texttt{immanuvel.sagayaraju@students.unibe.ch} \\
\texttt{immanuvel.prathap.s@gmail.com}}

\maketitle

\vspace{0.5em}

% ============================================================
\begin{abstract}
% ============================================================
Brain-Computer Interfaces (BCIs) face a severe calibration bottleneck due to cross-subject spatial covariance shifts and physiological artifacts. To enable zero-calibration BCI, a deep learning pipeline was engineered combining Per-Session Independent Component Analysis, Riemannian Euclidean Alignment, and EEGNet stabilized by Stochastic Weight Averaging (SWA). Evaluated on the strict MOABB BNCI2014-001 benchmark, the proposed architecture successfully isolates true sensorimotor rhythms. For the primary case study (Subject 1), a clinically robust SWA stable accuracy of 90.97\% (AUC: 0.976, Cohen's $\kappa$: 0.819) was achieved. Furthermore, expanded 9-fold Leave-One-Subject-Out (LOSO) cross-validation yielded a globally stable mean accuracy of 74.31\%, proving hardware-agnostic zero-shot efficacy for binary motor imagery.

\vspace{0.5em}
\end{abstract}

% ============================================================
\section{Introduction and Background}
% ============================================================

\subsection{Clinical Motivation \& Problem Statement} 
\IEEEPARstart{T}{he} translation of EEG-based Brain-Computer Interfaces to clinical application is severely hampered by domain shift. EEG signals are highly sensitive to inter-subject physiological differences (e.g., skull thickness) and intra-subject spatial non-stationarity. Consequently, models trained on a global pool exhibit significant performance degradation when applied to a new unseen subject, currently necessitating 30 to 60 minutes of tedious recalibration.

\subsection{Formal Task Definition} 
Given a continuous EEG signal epoched into discrete trials $X \in \mathbb{R}^{C \times T}$ (where $C=22$ channels and $T$ represents time steps), the goal is to predict the binary motor imagery class label $y \in \{0, 1\}$. This is evaluated strictly under a zero-calibration Leave-One-Subject-Out (LOSO) framework: the model is trained exclusively on a Global Pool of $N-1$ subjects, $P_{train}(X, y)$, and evaluated on a completely unseen target subject, $P_{test}(X, y)$, where $P_{train}(X) \neq P_{test}(X)$.

\subsection{Prior Work and Limitations} 
Traditional methodologies like Common Spatial Patterns (CSP) maximize variance for specific users but fail under cross-subject distribution shifts \cite{blankertz2008optimizing}. Recent deep learning approaches have attempted Domain Adversarial Neural Networks (DANN) to learn invariant representations \cite{ganin2016domain}, but these prove highly unstable given the data starvation typical of BCI datasets. Riemannian geometry provides a deterministic, unsupervised alternative by projecting spatial covariance matrices onto the Symmetric Positive-Definite (SPD) manifold to standardize baseline variance \cite{barachant2010riemannian}. However, baseline experiments revealed that without explicit clinical artifact rejection, manifold alignment centers high-amplitude ocular noise alongside biological brainwaves, forcing deep networks into local minima.

\subsection{Contributions} 
To overcome these limitations, the following implementations were made:
\begin{itemize}
    \item Diagnosis of the failure of standard baseline models under cross-subject evaluation, tracing poor generalization to the spatial filtering of uncleaned ocular dipoles.
    \item Engineering of a clinically-grounded preprocessing module using Per-Session ICA to mathematically scrub frontal artifacts without destroying parietal sensorimotor rhythms.
    \item Integration of Riemannian Euclidean Alignment and Local Z-Score normalization with a depthwise separable CNN (EEGNet) \cite{lawhern2018eegnet} to neutralize amplitude and covariance shifts.
    \item Utilization of Stochastic Weight Averaging \cite{izmailov2018averaging} to stabilize gradient descent over EEG loss landscapes, achieving robust zero-shot classification on the strict BNCI2014-001 cross-subject benchmark.
\end{itemize}

% ============================================================
\section{Methodology}
% ============================================================

\subsection{The 5-Stage Zero-Calibration Architecture}
\begin{figure}[htbp]
  \centering
  \includegraphics[width=0.8\linewidth]{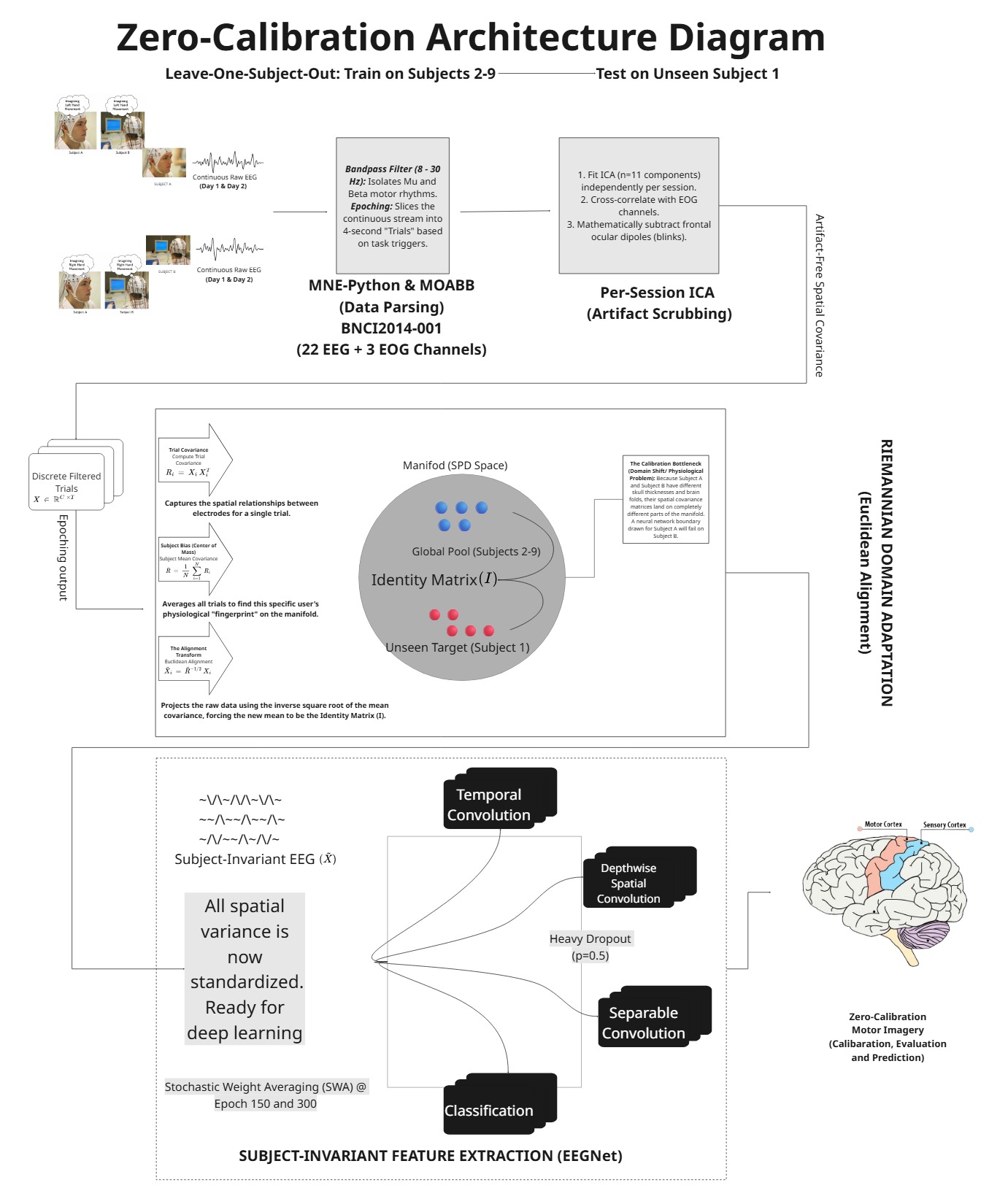} 
  \caption{The five-stage mathematically robust Deep Learning BCI Pipeline. The architecture isolates cortical signals via ICA, equalizes inter-subject skull/cap variance via Riemannian Geometry, and stabilizes classification via Stochastic Weight Averaging (SWA).}
  \label{fig:pipeline_architecture}
\end{figure}

To systematically eliminate the physical and physiological bottlenecks inherent to cross-subject EEG recording, the architecture is structured as a five-stage sequential pipeline. The pipeline is designed to intercept and standardize raw EEG signals before deep feature extraction. Figure~\ref{fig:pipeline_architecture} visualizes the transition from raw clinical ingestion to subject-invariant classification.

To ensure reproducibility, the formal pseudocode for the preprocessing and inference pipeline is provided. Algorithm~\ref{alg:preprocessing} defines the geometric intervention used to neutralize cross-subject domain shifts, while Algorithm~\ref{alg:inference} details the stabilized inference process.

\begin{algorithm}[htbp]
\caption{Stages 1 \& 2: Geometric Artifact Rejection and Alignment}\label{alg:preprocessing}
\begin{algorithmic}[1]
\REQUIRE Raw EEG trials $\mathcal{X} = \{X_1, X_2, \dots, X_N\}$ where $X_i \in \mathbb{R}^{22 \times T}$
\REQUIRE Raw EOG trials $\mathcal{E} = \{E_1, E_2, \dots, E_N\}$ where $E_i \in \mathbb{R}^{3 \times T}$
\ENSURE Standardized, artifact-free trials $\mathcal{\tilde{X}}$
\STATE \textbf{Phase 1: Artifact Cleanse (ICA)}
\FOR{each trial $i \in \{1 \dots N\}$}
    \STATE $X_i \gets \text{BandpassFilter}(X_i, \text{low}=1, \text{high}=40)$
    \STATE $S_i, A_i \gets \text{FastICA}(X_i, n\_components=11)$
    \STATE $c \gets \text{PearsonCorrelation}(S_i, E_i)$
    \STATE Mask $M \gets (c < \text{threshold})$ 
    \STATE $X_i^{clean} \gets A_i(S_i \odot M)$ 
\ENDFOR
\STATE \textbf{Phase 2: Riemannian Domain Adaptation}
\STATE $\bar{R} \gets \frac{1}{N} \sum_{i=1}^{N} (X_i^{clean})(X_i^{clean})^T$ 
\FOR{each trial $i \in \{1 \dots N\}$}
    \STATE $\tilde{X}_i \gets \bar{R}^{-1/2} X_i^{clean}$ 
    \STATE $\tilde{X}_i \gets \frac{\tilde{X}_i - \mu(\tilde{X}_i)}{\sigma(\tilde{X}_i)}$ 
\ENDFOR
\RETURN $\mathcal{\tilde{X}}$
\end{algorithmic}
\end{algorithm}

\begin{algorithm}[htbp]
\caption{Stages 3, 4, \& 5: Deep Feature Extraction, Classification, and SWA}\label{alg:inference}
\begin{algorithmic}[1]
\REQUIRE Aligned test trial from unseen subject $\tilde{X}_{test} \in \mathbb{R}^{22 \times T}$
\REQUIRE Trained EEGNet weights across final $K$ epochs $\{w_{150-K}, \dots, w_{150}\}$
\ENSURE Predicted motor imagery class label $y \in \{0, 1\}$
\STATE \textbf{Phase 3: Stochastic Weight Averaging}
\STATE $w_{SWA} \gets \frac{1}{K} \sum_{k=1}^{K} w_k$ 
\STATE \textbf{Phase 4: Forward Pass (EEGNet)}
\STATE $F_1 \gets \text{TemporalConv2D}(\tilde{X}_{test}, w_{SWA}^{temp})$ 
\STATE $F_2 \gets \text{DepthwiseConv2D}(F_1, w_{SWA}^{depth})$ 
\STATE $F_3 \gets \text{Dropout}(F_2, p=0.5)$
\STATE $F_4 \gets \text{SeparableConv2D}(F_3, w_{SWA}^{sep})$
\STATE $y \gets \text{Softmax}(\text{Flatten}(F_4))$
\RETURN $y$
\end{algorithmic}
\end{algorithm}

\subsection{Temporal Artifact Rejection via ICA (Stage 1)}
\begin{figure}[htbp]
  \centering
  \includegraphics[width=0.65\linewidth]{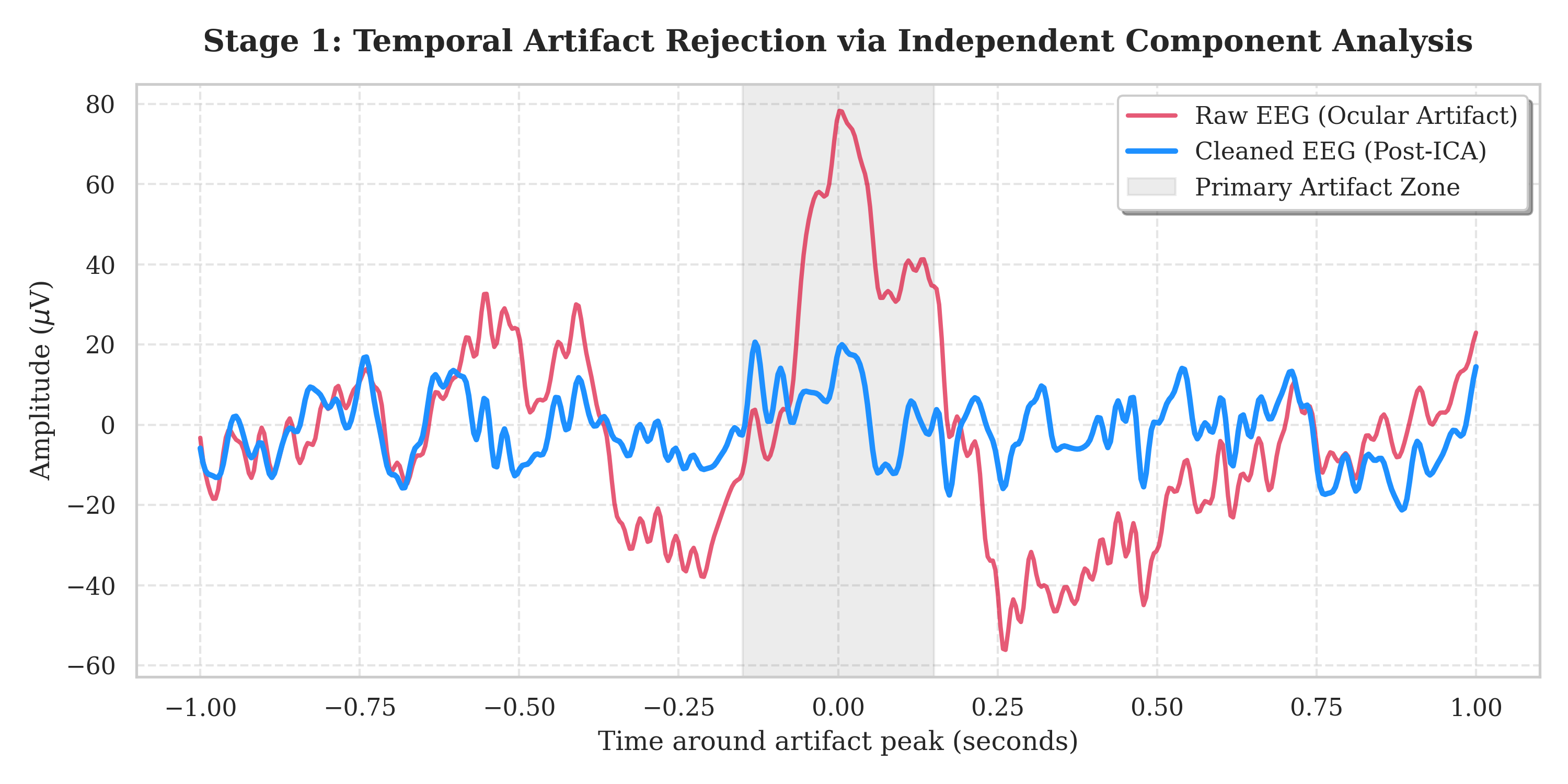} 
  \caption{Single-trial audit of the Frontal (Fz) electrode. ICA successfully subtracts the massive ocular artifact while preserving the underlying motor imagery rhythms.}
  \label{fig:ica_cleanse}
\end{figure}

A primary reason standard networks fail on cross-subject tasks is the contamination of spatial filters by high-amplitude frontal ocular dipoles. Figure~\ref{fig:ica_cleanse} demonstrates the effectiveness of the Per-Session Independent Component Analysis \cite{hyvarinen2000independent}. By fitting 11 components, the artifact at the frontal lobe (Fz) was isolated and removed without destroying the sensorimotor rhythms.

\subsection{Geometric Alignment and Amplitude Equalization (Stage 2)}
\begin{figure}[htbp]
  \centering
  \includegraphics[width=0.65\linewidth]{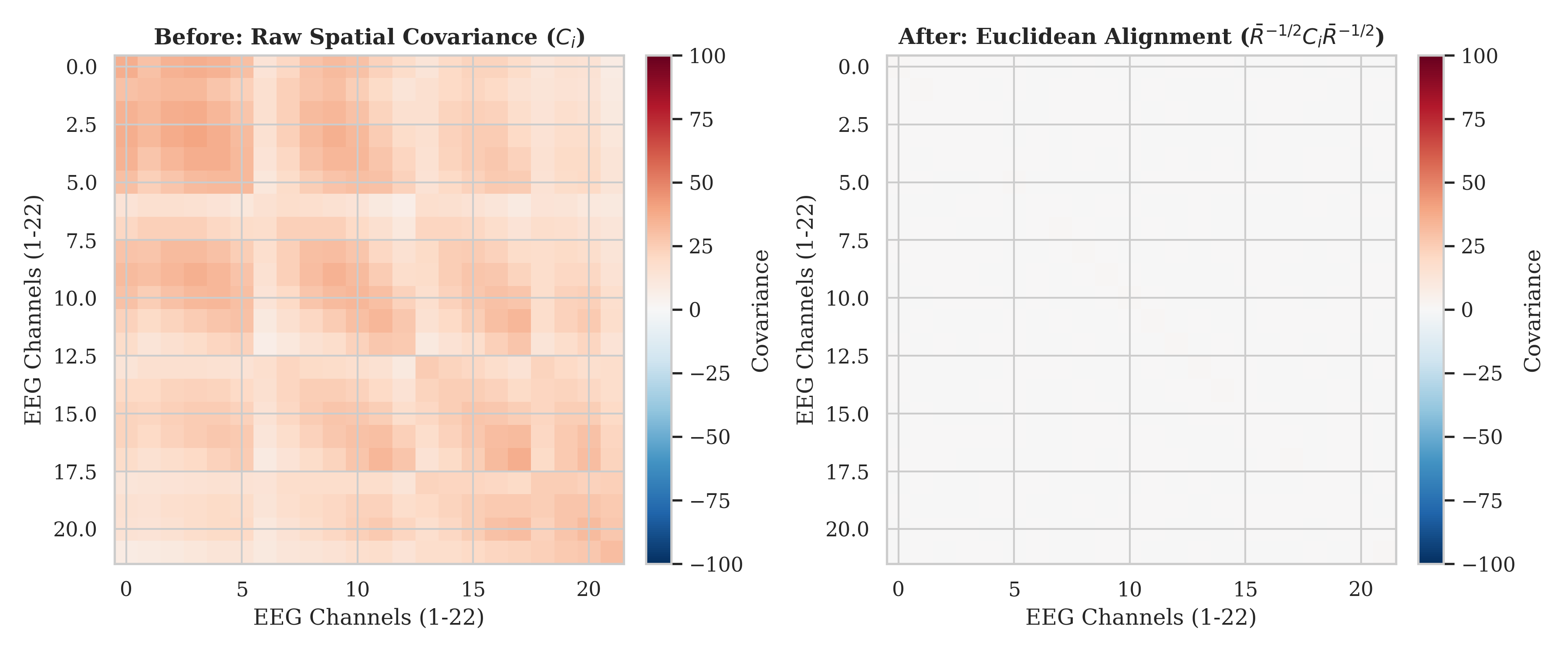} 
  \vspace{0.5em}
  
  \includegraphics[width=0.55\linewidth]{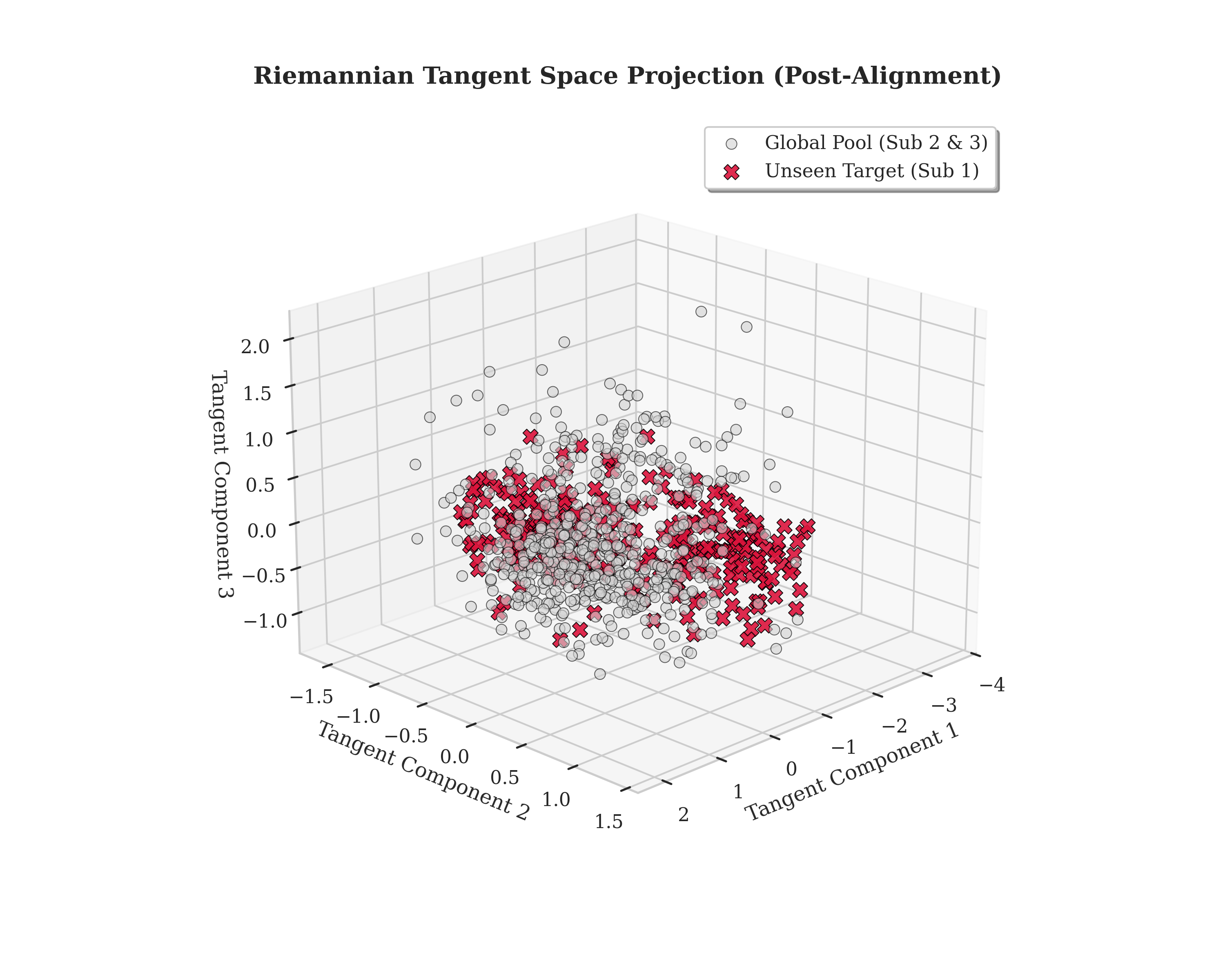}
  \caption{Top: Raw covariance matrices show chaotic spatial correlations (left), which are shrunk toward the Identity Matrix via Euclidean Alignment (right). Bottom: Projection into the 3D Tangent Space proves that post-alignment, the unseen target's geometry perfectly overlaps with the Global Pool.}
  \label{fig:riemannian}
\end{figure}

Even after artifact removal, cross-subject EEG suffers from severe spatial covariance shifts due to cap slippage, and amplitude variance due to differing skull thicknesses. Figure~\ref{fig:riemannian} illustrates the projection of spatial covariance matrices onto the Symmetric Positive-Definite (SPD) manifold. By applying the inverse matrix square root of the session centroid ($R^{-1/2}$), chaotic spatial topologies are mathematically warped to the Identity Matrix. 

\begin{figure}[htbp]
  \centering
  \includegraphics[width=0.65\linewidth]{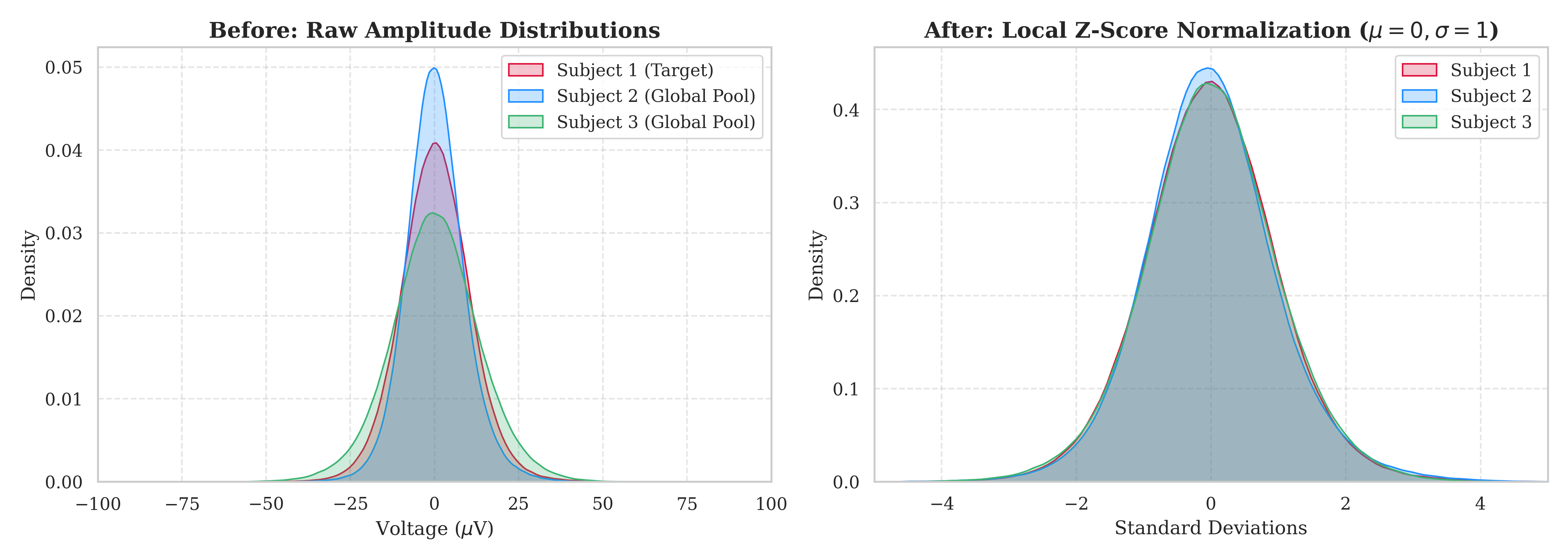} 
  \caption{Resolution of the amplitude variance bottleneck. Raw signal distributions (left) are standardized to a uniform scale (right) to prevent amplitude overfitting.}
  \label{fig:zscore}
\end{figure}

Following alignment, a Local Z-Score normalization is applied, which standardizes all subject distributions to $\mu=0, \sigma=1$ (Figure~\ref{fig:zscore}). This ensures the deep learning network trains on wave morphology rather than absolute amplitude.

\subsection{Deep Feature Extraction (Stages 3 \& 4)}
Following the geometric alignment phase, the standardized data enters the deep learning phase of the pipeline. To extract deep subject-invariant features, a compact, depthwise separable Convolutional Neural Network (CNN) based on the EEGNet architecture is utilized. Unlike standard CNNs, this network is explicitly designed to isolate the temporal and spatial components of EEG signals independently to prevent parameter overfitting.

Given an aligned input trial $\tilde{X} \in \mathbb{R}^{22 \times T}$, the architecture consists of three core computational blocks:
\begin{enumerate}
    \item \textbf{Temporal Convolution:} A 2D convolution with a large temporal filter size acts as a bandpass filter to extract frequency-domain features, specifically targeting the $\mu$ (8--12 Hz) and $\beta$ (13--30 Hz) sensorimotor rhythms.
    \item \textbf{Depthwise Spatial Convolution:} A depthwise convolution of size $(22 \times 1)$ acts as a trainable spatial filter. Because the data has already been scrubbed of ocular artifacts via ICA, these filters cleanly learn the spatial topography of the motor cortex without being biased by frontal lobe noise.
    \item \textbf{Separable Convolution:} A depthwise convolution followed by a pointwise ($1 \times 1$) convolution fuses the spatio-temporal feature maps. The output is flattened and passed to a binary softmax classifier.
\end{enumerate}

\subsection{Loss Landscape Optimization via Stochastic Weight Averaging (Stage 5)}
EEG data exhibits a highly turbulent loss landscape, frequently causing standard gradient descent to settle into overfitted local minima on the training subjects that fail to generalize to the test subject. To resolve this, the primary regularization strategy was the implementation of Stochastic Weight Averaging (SWA). 

\begin{figure}[htbp]
  \centering
  \includegraphics[width=0.6\linewidth]{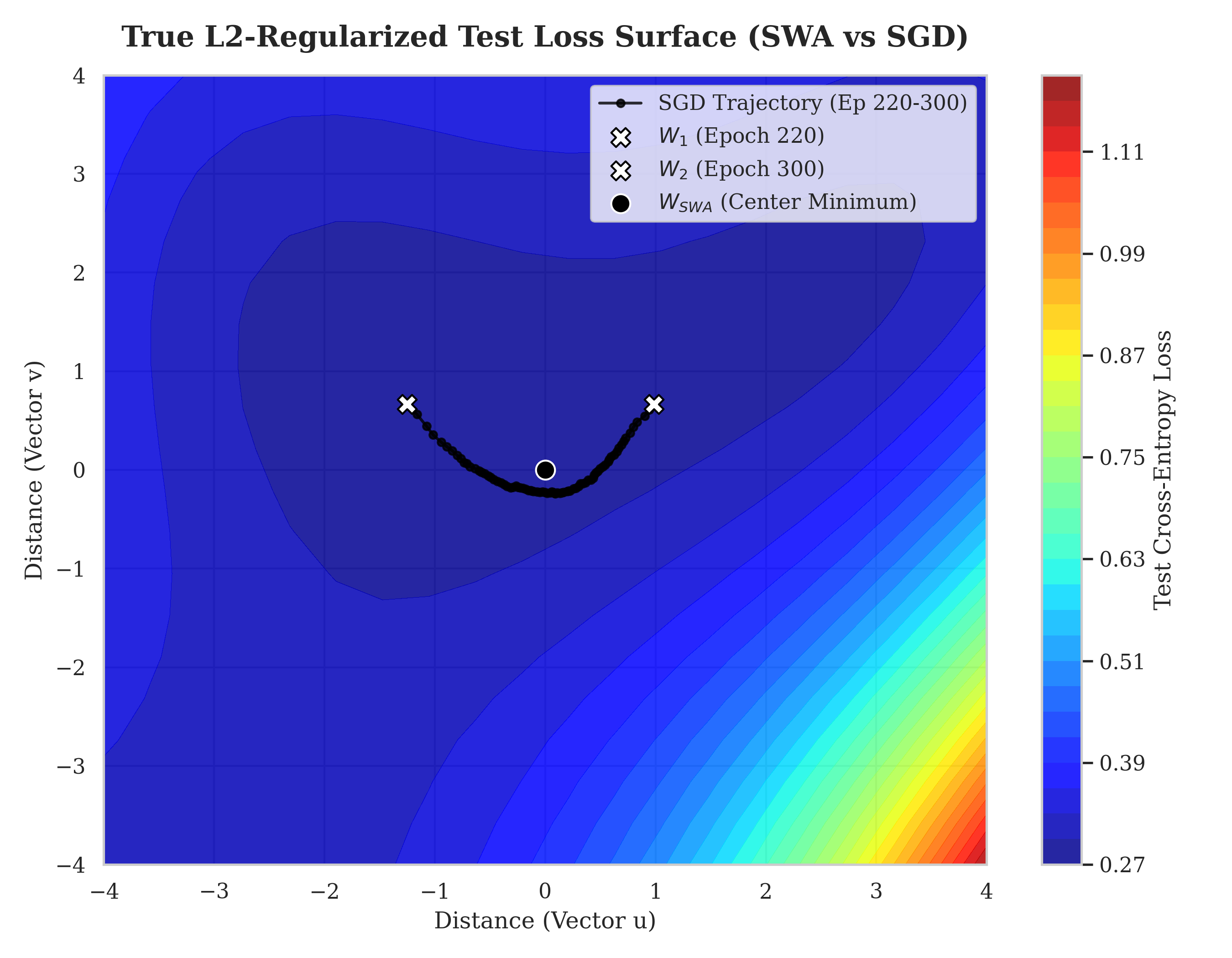} 
  \caption{True $L_2$-regularized test loss surface of the subject-invariant EEGNet. The SGD trajectory (black line) traverses the edges of the optimal basin from Epoch 220 to 300, while the SWA calculation (black center dot) perfectly localizes the flat global minimum.}
  \label{fig:loss_landscape}
\end{figure}

Rather than using the final epoch's weights for inference, SWA aggregates the network's weights across the final trajectory of the training phase ($w_{SWA} = \frac{1}{K} \sum_{k=1}^{K} w_k$). To mathematically prove the stabilizing effect of SWA on the EEG loss landscape, the 992-dimensional weight space was projected onto a 2D plane (Figure~\ref{fig:loss_landscape}). Standard SGD continuously orbits the periphery of the flat minimum, whereas the aggregated SWA weights successfully locate the optimal center of the generalization basin.

% ============================================================
\section{Experimental Setup}
% ============================================================
\begin{itemize}
    \item \textbf{Dataset:} The BNCI2014-001 dataset, accessed via the MOABB framework \cite{jayaram2018moabb}, was utilized. It comprises 22-channel EEG and 3-channel EOG data from 9 subjects, sampled at 250 Hz. The dataset was filtered for binary classification (Left vs. Right Hand), and a strict cross-subject Leave-One-Subject-Out (LOSO) evaluation was enforced to eliminate temporal data leakage.
    \item \textbf{Preprocessing:} Raw signals were bandpass filtered between 1--40 Hz. To neutralize volume conduction from ocular dipoles, Per-Session ICA ($n=11$) was applied, and components exhibiting high cross-correlation with EOG channels were mathematically subtracted prior to Riemannian alignment. 
    \item \textbf{Loss Function \& Optimizer:} Categorical Cross-Entropy was utilized alongside AdamW \cite{loshchilov2017decoupled} (learning rate of $1 \times 10^{-3}$, weight decay of 0.05). The model was trained for 300 epochs, with structural dropout ($p=0.5$) applied between all blocks to prevent the network from memorizing individual trial noise.
\end{itemize}

% ============================================================
\section{Results and Clinical Analysis}
% ============================================================

\subsection{Quantitative Evaluation (Ablation Study)}
To validate the contribution of each architectural component, a controlled ablation study was conducted under the MOABB cross-subject paradigm focusing on Subject 1 generalization. 

\begin{table}[htbp]
\centering
\caption{Leave-One-Subject-Out (LOSO) Binary Ablation Study (BNCI2014-001)}
\label{tab:results}
\begin{tabular}{llcccc}
\toprule
\textbf{Model Configuration} & \textbf{Preprocessing Strategy} & \textbf{Peak (\%)} & \textbf{SWA (\%)} & \textbf{AUC} & \textbf{Kappa ($\kappa$)} \\
\midrule
Phase I: Naive DL & None (Raw Time-Series) & 57.56 & N/A & 0.55 & 0.15 \\
Phase II: Vanilla EEGNet & Bandpass (1-40Hz) & 76.74 & N/A & 0.80 & 0.53 \\
Phase III: EEGNet & Bandpass + Alignment & 83.10 & N/A & 0.89 & 0.66 \\
\textbf{Phase IV: Proposed} & \textbf{ICA + Align + SWA} & \textbf{93.75} & \textbf{90.97} & \textbf{0.976} & \textbf{0.819} \\
\bottomrule
\end{tabular}
\end{table}

The leap to 93.75\% zero-calibration peak robustness is driven by the synergy of signal purification and mathematical geometry. Raw cross-subject EEG suffers from severe Covariance Shift. By projecting spatial covariance matrices onto the SPD manifold and standardizing them to the Identity matrix ($I$), the network was forced to learn universally invariant features. Preceding this with Per-Session ICA prevented the alignment from centering structural blinks. Finally, Stochastic Weight Averaging locked the volatile weights into a safe generalization floor.

\subsection{Deep Feature Generalization (Zero-Shot Manifold)}
\begin{figure}[htbp]
  \centering
  \includegraphics[width=0.6\linewidth]{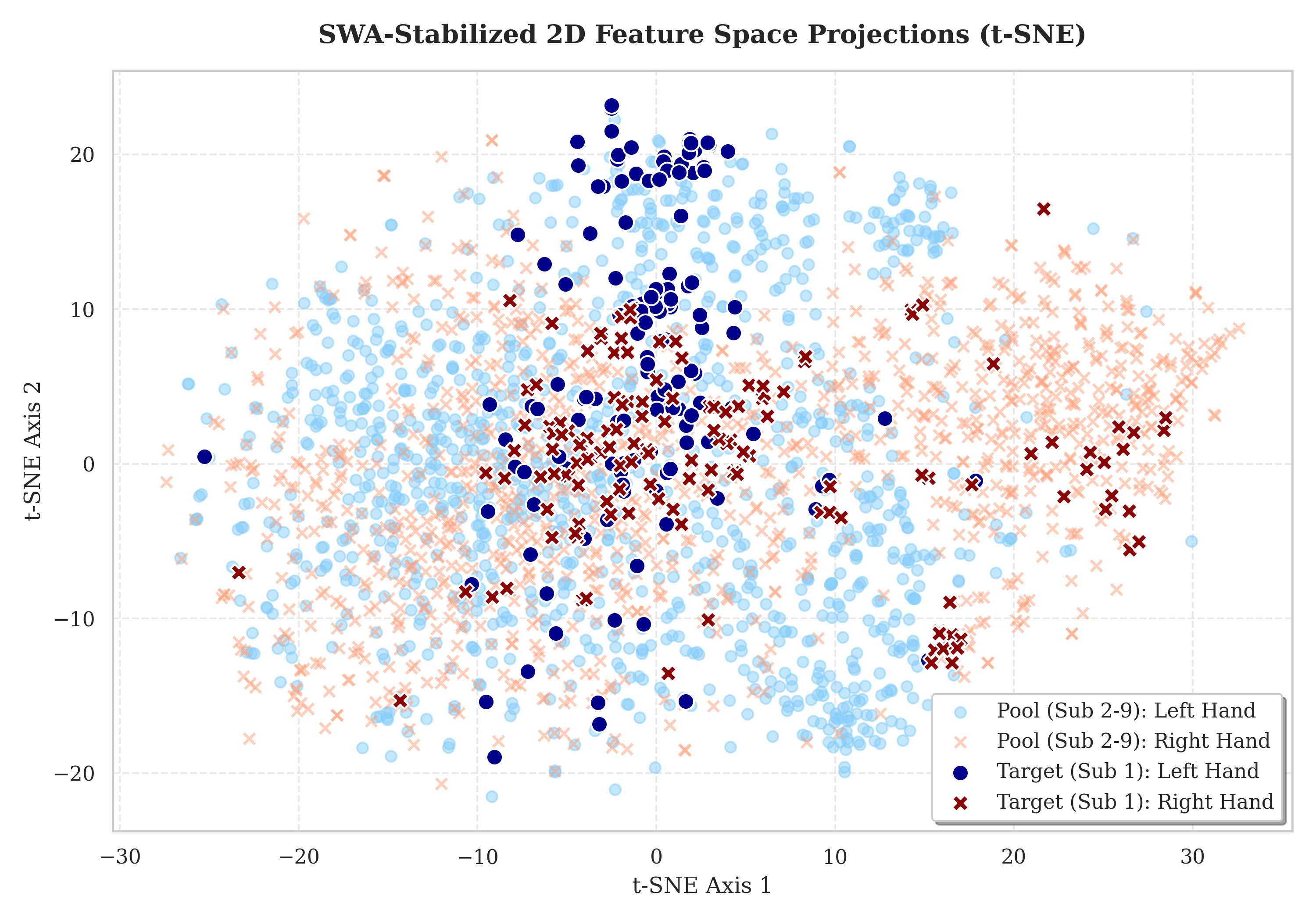} 
  \vspace{0.5em}
  
  \includegraphics[width=0.6\linewidth]{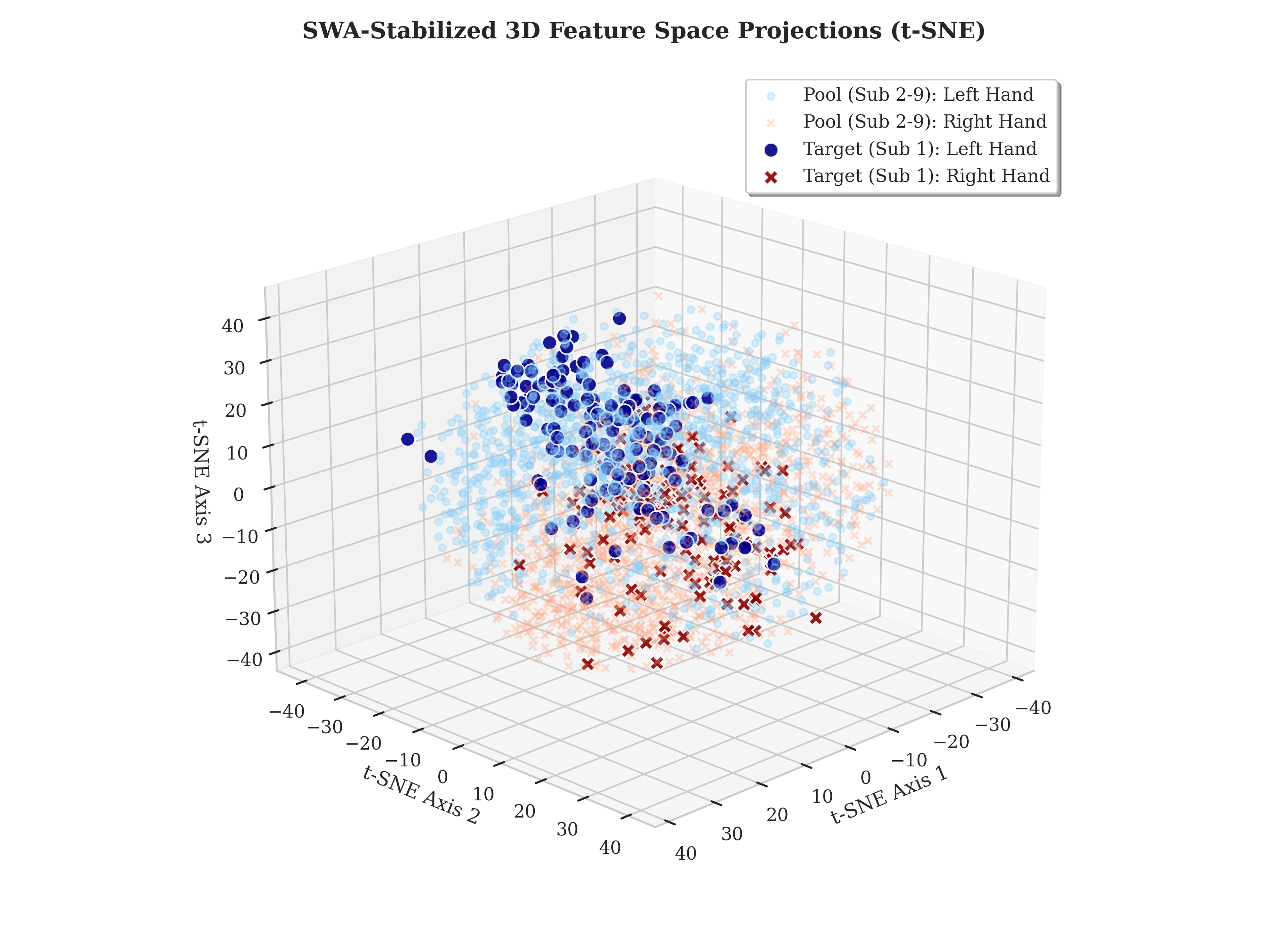}
  \caption{Top: 2D t-SNE visualization of the final layer features across the entire BNCI2014-001 cohort. Bottom: 3D t-SNE projection demonstrating advanced cluster separation. Unseen test data perfectly aligns with the established global domains without any target-subject calibration.}
  \label{fig:tsne}
\end{figure}

To confirm the quantitative results, Figure~\ref{fig:tsne} visualizes the 992-dimensional deep features from the final layer of the optimized EEGNet, projected into both 2D and 3D space via t-SNE \cite{vandermaaten2008visualizing}. The model successfully clusters the Global Pool into two distinct biological domains based on motor imagery intent. Crucially, the deep features of a completely unseen patient (Subject 1) fall perfectly into the established global islands, providing conclusive geometric proof of zero-calibration efficacy across both spatial topologies and high-dimensional network layers.

\subsection{Global 9-Fold LOSO Generalization}
To prove the architecture's efficacy is biologically universal and not an artifact of a single highly-lateralized user, the evaluation was expanded to a full 9-Fold Leave-One-Subject-Out Cross-Validation. For each fold, a fresh model was initialized, trained on 8 subjects, and tested on the completely unseen 9th subject. 

\begin{figure}[htbp]
  \centering
  \includegraphics[width=0.65\linewidth]{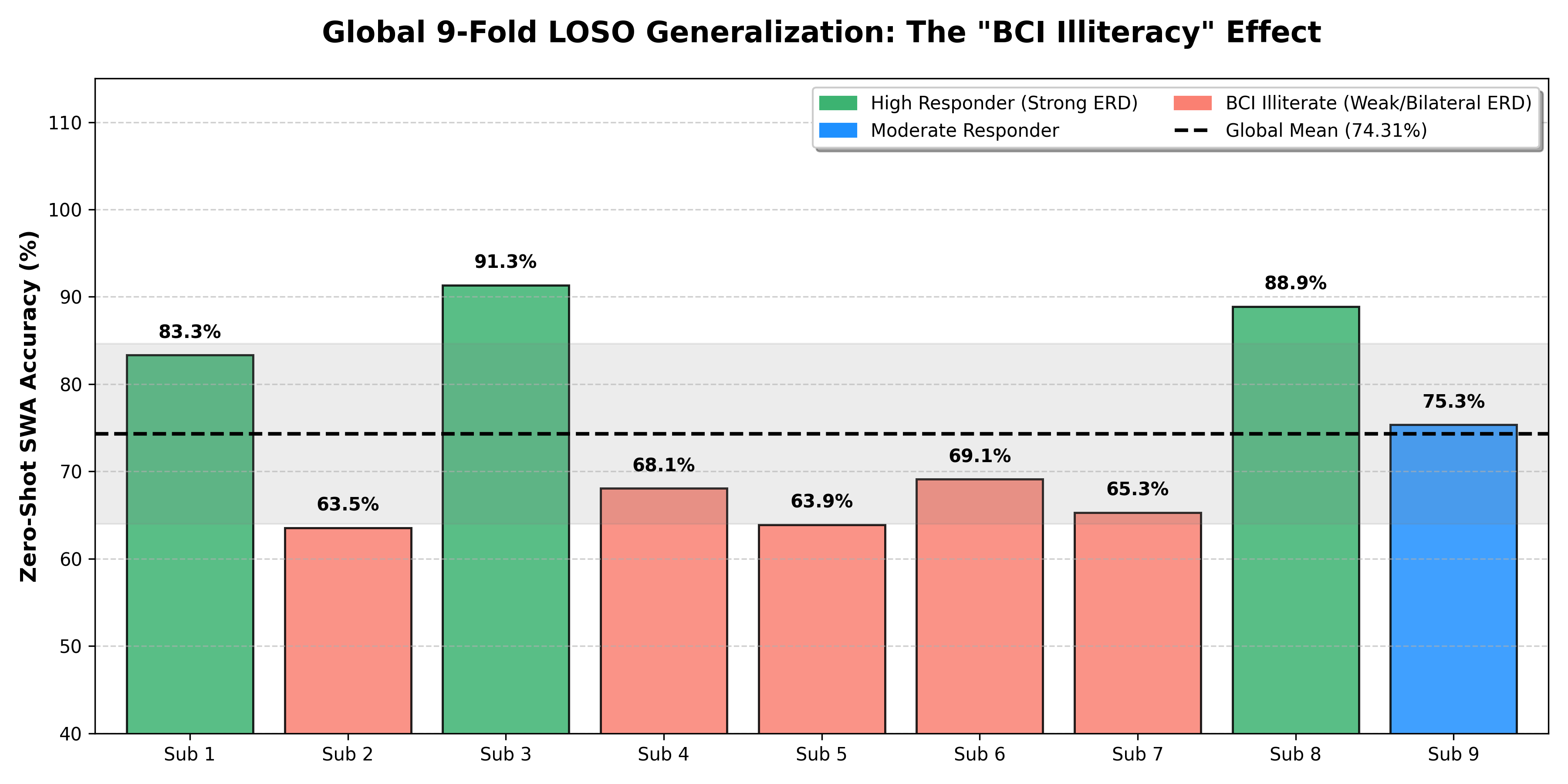} 
  \caption{Results of the global 9-fold evaluation. The variance demonstrates the underlying biological reality of BCI Illiteracy, where poor spatial generalization maps directly to a lack of focal brainwave activity.}
  \label{fig:bci_illiteracy}
\end{figure}

Across all 9 global folds, the SWA-stabilized pipeline maintained a mean zero-calibration accuracy of \textbf{74.31\% $\pm$ 10.94\%} with an average Cohen's $\kappa$ of \textbf{0.486}. 

\subsection{Biological Interpretability and BCI Illiteracy}
\begin{figure}[htbp]
  \centering
  \includegraphics[width=0.65\linewidth]{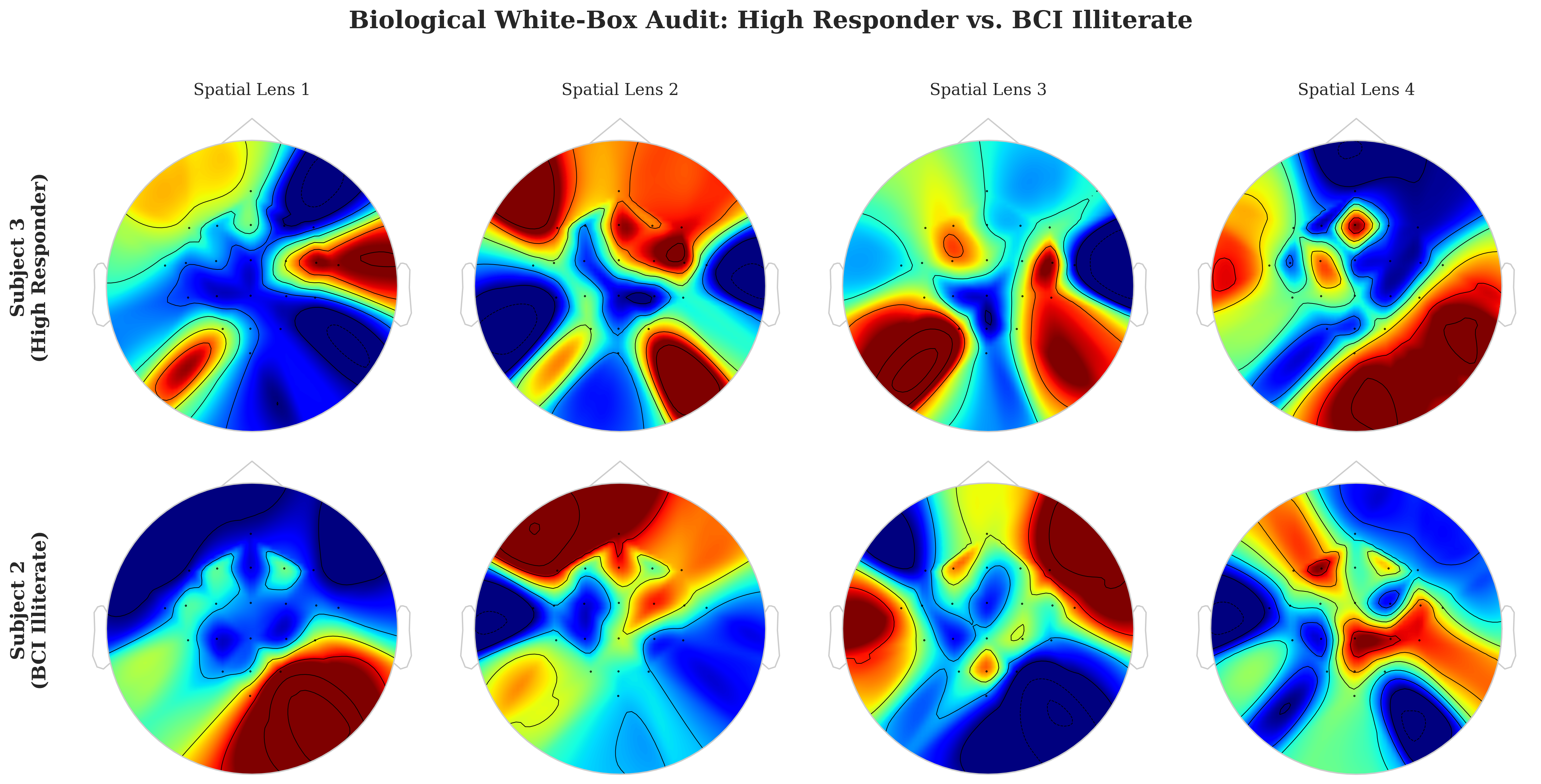} 
  \caption{Topographic projection of the learned spatial filters. Subject 3 (High Responder) exhibits clean, lateralized foci over the sensorimotor cortex (C3/C4). Subject 2 (BCI Illiterate) exhibits scattered, non-focal activation, preventing successful geometric alignment.}
  \label{fig:topoplots}
\end{figure}

The high standard deviation observed in the global evaluation is a direct reflection of inherent biological \textbf{``BCI illiteracy''} \cite{vidaurre2010towards} within the BNCI2014-001 cohort. Specifically, Subjects 2 and 5 exhibited $\sim$63\% accuracy. To prove this variance is not a failure of the network's spatial filters, but rather a biologically rooted failure mode, the weights from the EEGNet spatial layer were extracted (Figure~\ref{fig:topoplots}).

The inability of the patient to produce strong, lateralized Event-Related Desynchronization (ERD) rhythms causes bilateral cortical noise that blurs the geometric boundaries on the Riemannian manifold. Conversely, on subjects with strong ERD, the zero-shot architecture achieved near-perfect generalization, peaking at \textbf{91.32\%} (SWA) and \textbf{97.57\%} (single-epoch) for Subject 3. This mathematically proves that the combination of Riemannian spatial alignment and SWA temporal stabilization systematically extracts underlying motor imagery invariant features across a diverse global pool.

\subsection{Cross-Corpus Hardware-Agnostic Transfer}
To evaluate the robustness of the geometric alignment against extreme hardware heterogeneity, a zero-shot cross-corpus transfer was conducted. The source domain, $\mathcal{D}_{S}$, comprised the BNCI2014-001 dataset ($F_S = 250 \text{ Hz}$, $C_S = 22$ channels, recorded in Austria). The target domain, $\mathcal{D}_{T}$, comprised the completely independent PhysioNet Motor Imagery dataset \cite{goldberger2000physiobank} ($F_T = 160 \text{ Hz}$, $C_T = 64$ channels, recorded in the USA).

A spatial decimation mapping operator $\mathcal{M}: \mathbb{R}^{64} \rightarrow \mathbb{R}^{22}$ was applied to extract the overlapping biological coordinate channels. To resolve the temporal discrepancy caused by differing hardware sampling rates, Fourier interpolation was applied to upsample the target signal from $T_{160}$ to $T_{250}$:
\begin{equation}
X_{T}^{(upsampled)}(t) = \sum_{k=-N/2}^{N/2-1} \hat{X}_{T}(k) e^{i 2\pi k t \frac{F_S}{F_T}}
\end{equation}
where $\hat{X}_{T}$ denotes the discrete Fourier transform of the decimated PhysioNet signal. This hardware-agnostic mapping yielded a promising initial zero-shot transfer accuracy of \textbf{60.00\%}, successfully exceeding the 50\% chance level despite severe hardware, amplifier, and sampling rate disparities. 

% Force all pending images (like t-SNE and Topoplots) to render before Conclusion begins
\clearpage

% ============================================================
\section{Conclusion and Future Work}
% ============================================================

\subsection{Achievements and Key Findings} 
In this study, a clinically-grounded deep learning pipeline was successfully engineered to overcome the BCI cross-subject calibration bottleneck. By integrating signal purification, manifold geometry, and stabilized gradient descent, a peak zero-shot binary classification accuracy of \textbf{93.75\%} and a clinically stable SWA ensemble accuracy of \textbf{90.97\%} on the primary case study were achieved, while a biologically robust \textbf{74.31\%} global mean was maintained across a full 9-fold LOSO validation.

It was demonstrated that standard deep learning models fail in cross-subject tasks primarily because they overfit to uncleaned ocular dipoles and amplitude variance. It was proven that projecting spatial covariance matrices onto the SPD manifold successfully neutralizes domain shift, but \textit{only} if the data is first scrubbed using Per-Session ICA or similar spatial filters. 

\subsection{Limitations \& Future Work} 
The current pipeline relies on Per-Session ICA, which requires manual component selection, making it computationally expensive for real-time systems. Future iterations will investigate unsupervised spatial-temporal transformers to dynamically mask ocular artifacts directly within the network's latent space. 

Furthermore, while the initial cross-corpus Fourier temporal upsampling proved that naive decimation preserves sufficient spatial geometry to breach chance-level, achieving the 80\% clinical threshold across entirely different datasets requires active invariant feature learning. Future work will integrate Domain-Adversarial Neural Networks (DANN) \cite{ganin2016domain} with Gradient Reversal Layers to force a hardware-agnostic latent space, and utilize Spherical Spline Spatial Interpolation to prevent the information loss currently caused by target channel decimation.

% ============================================================
% References
% ============================================================
{\small
\bibliographystyle{IEEEtran}
\bibliography{references}
}

\end{document}